\documentclass[twocolumn,floatfix,prl,showpacs,superscriptaddress]{revtex4}
\usepackage{graphicx}
\usepackage{amsmath}

\begin{document}
	\title{Unconventional spin dynamics  in the honeycomb-lattice material $\alpha$-RuCl$_3$: high-field ESR studies}

\author{A.~N.~Ponomaryov}
\affiliation{Dresden High Magnetic Field Laboratory (HLD-EMFL), Helmholtz-Zentrum Dresden-Rossendorf, D-01328 Dresden, Germany}
\author{E.~Schulze}
\affiliation{Dresden High Magnetic Field Laboratory (HLD-EMFL), Helmholtz-Zentrum Dresden-Rossendorf, D-01328 Dresden, Germany}
\affiliation{Institut f\"{u}r Festk\"{o}rperphysik und Materialphysik, TU Dresden, 01062 Dresden, Germany}
\author{J.~Wosnitza}
\affiliation{Dresden High Magnetic Field Laboratory (HLD-EMFL), Helmholtz-Zentrum Dresden-Rossendorf, D-01328 Dresden, Germany}
\affiliation{Institut f\"{u}r Festk\"{o}rperphysik und Materialphysik, TU Dresden, 01062 Dresden, Germany}
\author{P.~Lampen-Kelley}
\affiliation{Materials Science and Technology Division, Oak Ridge National Laboratory, Oak Ridge, TN 37821, USA}
\affiliation{Department of Materials Science and Engineering, University of Tennessee, Knoxville, TN 37821, USA}
\author{A.~Banerjee}
\affiliation{Quantum Condensed Matter Division, Oak Ridge National Laboratory, Oak Ridge, TN 37821, USA}
\author{J.-Q.~Yan}
\affiliation{Materials Science and Technology Division, Oak Ridge National Laboratory, Oak Ridge, TN 37821, USA}
\author{C.~A.~Bridges}
\affiliation{Chemical Science Division, Oak Ridge National Laboratory, Oak Ridge, TN 37821, USA}
\author{D.~G. Mandrus}
\affiliation{Materials Science and Technology Division, Oak Ridge National Laboratory, Oak Ridge, TN 37821, USA}
\author{S.~E.~Nagler}
\affiliation{Quantum Condensed Matter Division, Oak Ridge National Laboratory, Oak Ridge, TN 37821, USA}
\author{A.~K.~Kolezhuk}
\affiliation{Institute of High Technologies, Taras Shevchenko National
University of Kyiv, 03022 Kyiv, Ukraine}
\affiliation{Institute of Magnetism, National Academy of Sciences and
  Ministry of Education, 03142 Kyiv, Ukraine}
\author{S.~A.~Zvyagin}
\affiliation{Dresden High Magnetic Field Laboratory (HLD-EMFL), Helmholtz-Zentrum Dresden-Rossendorf, D-01328 Dresden, Germany}
	
\date{\today}
	
\begin{abstract}

We present high-field electron spin resonance (ESR) studies of the honeycomb-lattice material 
 $\alpha$-RuCl$_3$, a prime candidate to
exhibit Kitaev physics. Two  modes of antiferromagnetic resonance were detected in the  zigzag ordered phase,  with magnetic field applied in the $ab$ plane. 
A very rich excitation spectrum was observed in the field-induced quantum paramagnetic phase.   The obtained data are compared with  results of  recent numerical calculations,  strongly  suggesting a very unconventional    multiparticle character  of the spin dynamics in  $\alpha$-RuCl$_3$.  The  frequency-field diagram  of the lowest-energy ESR mode is found consistent with  the behavior of the  field-induced energy gap, revealed by thermodynamic measurements.


\end{abstract}
	
\maketitle
	
Spin systems with honeycomb structures have recently attracted a great
deal of attention, both theoretically and experimentally. It was
proposed that some of such systems can be  experimental realizations of
the Kitaev-Heisenberg model \cite{Kitaev}, which encompasses a variety
of possible magnetic ground states (from a conventional N\'{e}el order
to a  quantum spin liquid) and emergent fractional excitations
(e.g., Majorana fermions and gauge fluxes)
\cite{Chaloupka,Yadav,Petrova,Baskaran,Knolle,Janssen,Do}.  An essential
peculiarity of this model is the presence of anisotropic
bond-dependent interactions, defined in the Hamiltonian
(Eq. \ref{Ham}) by the Kitaev parameter $K$:

\begin{eqnarray}
\label{Ham} \mathcal{H}&=&J\sum_{\langle i,j\rangle}\pmb{S}_i\cdot\pmb{S}_j+ K\sum_{\langle i,j\rangle_m}\pmb{S}_i^m\cdot\pmb{S}_{j'}^m-\pmb{h}\sum_{i}\pmb{S}_i,
\end{eqnarray}
where $S_i$ and  $S_j$  are spin-1/2 operators at sites $i$ and  $j$,
respectively,    $J$ is the Heisenberg exchange interaction,
$m=x,~y,~z$ label the three different links of the lattice, and $h$ is
the uniform magnetic field. The Kitaev  physics is thought to be realized in Ir-based magnets   (such as $A_2$IrO$_3$, $A=$ Na or Li \cite{Chaloupka,Jackeli,Singh,Takayama}), where,  due to a strong spin-orbital interaction,  the multiorbital $5d$ $t_{2g}$ state can be mapped into a single orbital state with a pseudospin $j_{eff}$ = 1/2.

\begin{figure} [!h]

\begin{center}
\vspace{0mm}
\includegraphics[width=0.45\textwidth]{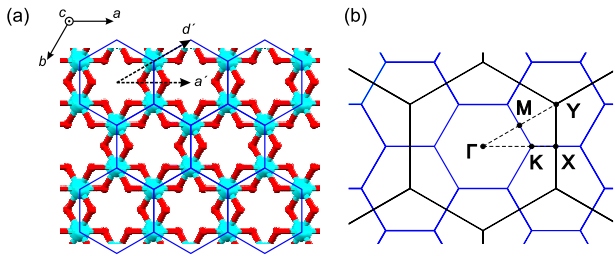}
\vspace{0mm}
\caption{\label{fig:STR}  (\textit{a}) Schematic view of the
  $\alpha$-RuCl$_3$ crystal structure in the trigonal setting  ($ab$ view).  The Ru$^{3+}$ ions are shown in cyan, while the
  Cl$^{-}$ ions are shown in red. (\textit{b}) The reciprocal lattice
  of the honeycomb lattice (first and second Brillouin zones are
  shown). }
\end{center}
\end{figure}

Recently, the honeycomb-lattice material $\alpha$-RuCl$_3$  [Fig.~\ref{fig:STR}(a)] has been proposed as another promising
candidate to exhibit Kitaev physics. The local cubic symmetry of
$\alpha$-RuCl$_3$ is almost perfect, in contrast to the iridates.  As
revealed experimentally \cite{Kubota,Johnson}, the magnetic
susceptibility and magnetization of $\alpha$-RuCl$_3$ are very
anisotropic, evident of the low-spin state of Ru$^{3+}$.
Low-temperature neutron scattering measurements \cite{Banerjee,Ran}
suggested  a collinear zigzag-ordered magnetic structure, which is one
of the magnetic states predicted by the Kitaev-Heisenberg model.
Magnetic field applied in the $ab$ plane suppresses the long-range
magnetic order, so that above the critical field
$H_c\approx 7$ T the system is in a  quantum paramagnetic  phase \cite{Kubota,Johnson,Leahy}. At about 23~T the system undergoes the transition into the magnetically saturated  phase \cite{Johnson,Rem_FSPS}. 
One exciting property of the quantum paramagnetic phase  is the presence of 
a field-induced energy gap, revealed experimentally  by means of nuclear magnetic resonance and heat-transport measurements \cite{Baek,Sears,Hentric,Wolter}.

In this work, we present results of systematic high-frequency electron spin resonance (ESR) studies of $\alpha$-RuCl$_3$ in
magnetic fields up to 16 T, allowing us to gain a deeper insight into the nature and peculiarities of the spin dynamics in this material 
across different phases of its phase diagram.

Single-crystal $\alpha$-RuCl$_3$ samples  with typical sizes  of 3x3x0.5 mm$^3$ were prepared using a vapor transport technique starting from pure RuCl$_3$ powder  \cite{Banerjee}. The samples were characterized using standard thermodynamic techniques;  the obtained specific-heat data are consistent with published results  \cite{Kubota}, exhibiting  a sharp peak at $T_N\simeq 7.5$ K and the onset of a broad anomaly near 14 K (the latter is attributed to the presence of stacking faults  \cite{Johnson}). The ESR  measurements were performed employing  a 16 T   transmission-type  ESR spectrometer,  similar to that  described in Ref. \cite{Zvyagin_INSR}. In our experiments, a set of backward-wave oscillators, Gunn diodes, and VDI microwave sources  were used, allowing us to probe magnetic excitations in this material in the very broad quasi-continuously covered frequency range from ca 50 GHz to 1.2 THz. The experiments were done in the Voigt configuration with magnetic field applied in the $ab$ plane at temperatures down to 1.4 K.

\begin{figure}[!h]

\includegraphics[width=0.47\textwidth]{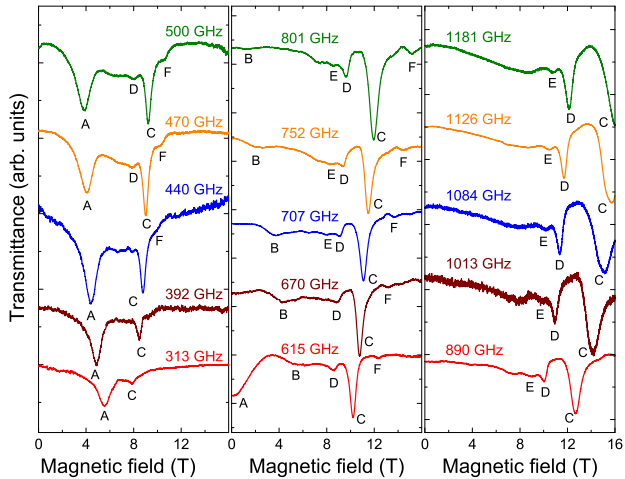}
\caption{\label{fig:SP}  Examples of ESR spectra taken at
  different frequencies ($H \parallel [110]$, $T=1.4$ K). The spectra are offset
  for clarity.}
\end{figure}

A very rich excitation spectrum was observed  at a temperature of 1.4 K (Fig.~\ref{fig:SP}), revealing  the presence of  six absorption lines:  modes A and B were detected  in the low-field zigzag-ordered phase, while modes C, D, E, and F  in the field-induced quantum paramagnetic phase.   

\begin{figure} [!h]

\includegraphics[width=0.5\textwidth]{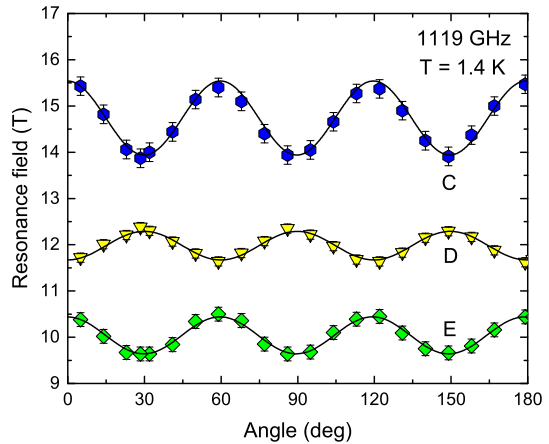}
\caption{\label{fig:AD}  Angular dependence of the ESR modes C, D, and E, 
  taken at a frequency of 1119 GHz, $T=1.4$ K. The angles  $0^\circ$ and $30^\circ$ correspond to [110] and [100] direction, respectively.   Lines are guides to the eye.}
\end{figure}

Angular dependences of resonance fields for  modes C, D, and E (measured   at a frequency of 1119 GHz, $T=1.4$ K) are shown in Fig.~\ref{fig:AD}.  The experiment reveals  the 60$^\circ$ periodicity of  ESR fields, as expected for a honeycomb structure. The angles  $0^\circ$ and $30^\circ$ correspond to  [110] and [100] direction, respectively.  We would like to stress the  importance of these measurements, allowing us  to confirm  the very high,  twin-free, quality of the single-crystalline samples we used.

\begin{figure}[!h]
\begin{center}
\includegraphics[width=0.55\textwidth]{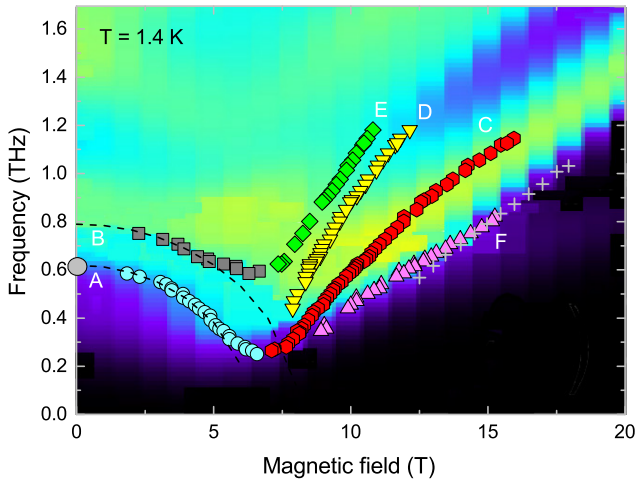}
\caption{\label{fig:FFD_b}  Frequency-field diagrams of ESR
  excitations in $\alpha$-RuCl$_3$ for $H \parallel [110]$, $T=1.4$ K (symbols). The experimental data are   shown together with the 
  ESR response at the $\Gamma$ point for $H \parallel a$, $h_{\omega}\perp H$ (color scale), obtained numerically by means of ED for clusters up to 24 spins \cite{Winter_2}. 
	Zero-field  excitations observed  at the $\Gamma$ point by means of neutron scattering \cite{Banerjee_2} and THz spectroscopy  \cite{Little}  measurements are shown by a gray circle. 
	Dash lines correspond to the fit results as described in \cite{rem_1}. The field-induced gap revealed  in the high-field
 phase by means of heat-transport  measurements \cite{Hentric} is denoted  by crosses. }
\end{center}
\end{figure}


\begin{figure} [!h]
\vspace{-1.5cm}
\includegraphics[width=0.5\textwidth]{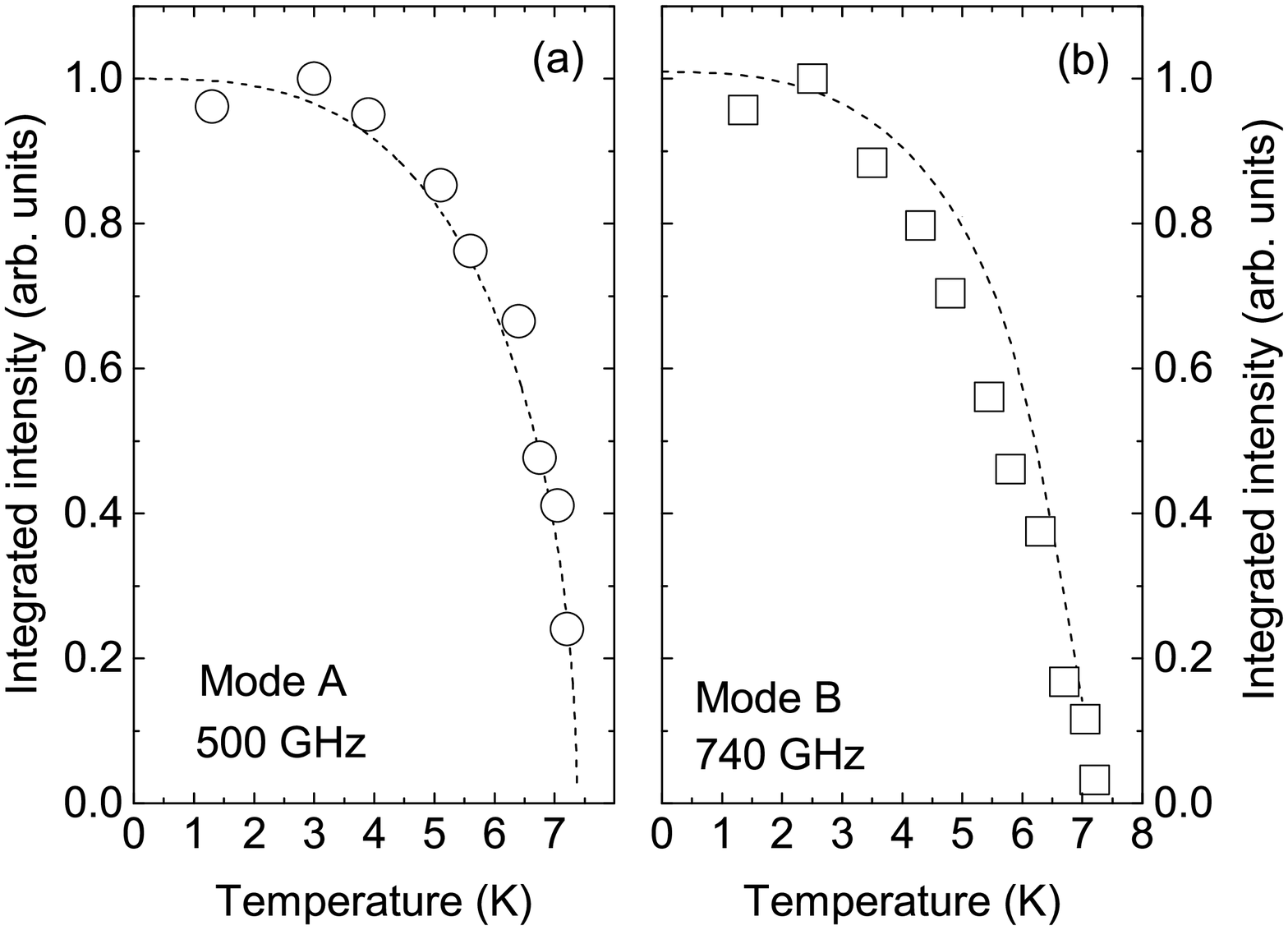}
\caption{\label{fig:AFMR_I} Temperature dependence of integrated  ESR intensities for modes A (a) and B (b), taken at frequencies 500 and 740 GHz, respectively ($H \parallel [100]$).  }
\end{figure}

The frequency-field diagrams of ESR excitations  for  $H \parallel [110]$ are shown in
Fig.~\ref{fig:FFD_b}. As mentioned, two gapped ESR modes, A and B, were observed below $H_c$, where
the system is in the  zigzag-ordered  state.   The intensities of both modes decrease with
increasing temperature, and  at about $T_N\simeq 7.5$ K both resonance lines  vanish (Fig.~\ref{fig:AFMR_I}), evidenced that the detected ESR modes are indeed the modes 
of  antiferromagnetic resonance (AFMR) in the long-range magnetically ordered zigzgag  phase.  
The extrapolation of the frequency-field dependences of modes  A and B  to zero field \cite{rem_1}
revealed gaps $\Delta_A=620$ GHz (which corresponds to 2.56 meV) and $\Delta_B=790$ GHz (3.27 meV),  respectively  [Fig.~\ref{fig:FFD_b}]. The gap  $\Delta_A$ GHz is consistent  with  results of inelastic neutron scattering at the $\Gamma$ point \cite{Banerjee_2}, time-domain THz spectroscopy \cite{Little} measurements  (shown in Fig.~\ref{fig:FFD_b} by a gray circle) and calculations \cite{Winter}, providing clear evidence of magnetic excitations at the center of the Brillouin zone.  It is important to mention that due to absence of     the inversion symmetry  on the second-nearest-neighbour bonds, the Dzyaloshinskii-Moria (DM) term in $\alpha$-RuCl$_3$ is allowed, allowing in its turn,   ESR  transitions at the K point (Fig.~\ref{fig:STR}). Such excitations at the Brillouin-zone boundaries   were observed in a number of  multisublattice  antiferromagnets (see, e.g., \cite{Bar,Sakai,Zhe_BSCO}) and known as $exchange$  modes. Thus,  we speculate that mode B can be interpreted as such an exchange AFMR mode. 
Both AFMR branches demonstrate pronounced softening with increasing field. It is noticeable, that the  lowest-energy observed AFMR  gap, corresponding to magnetic excitations at the $\Gamma$ point,  remains open   at $H_c$.  The presence of additional anisotropic terms (such as, e.g., the staggered DM interaction) can be a reason of the observed phenomenon \cite{Zvyagin_CUP}.

Now we would like to focus on the high-field spin dynamics in $\alpha$-RuCl$_3$.   Above $H_c\approx 7$ T,  the long-range zigzag order  is suppressed, and  the system is in the magnetically disordered but strongly correlated quantum paramagnetic  phase. There, four  modes were observed. The corresponding frequency-field diagrams of magnetic excitations  for $H \parallel [110]$  are shown   in  Fig.~\ref{fig:FFD_b}.  In our experiments, the most intensive mode C  can be detected  at temperatures up to $\sim 20$ K  (Fig.~\ref{fig:SP}). Mode D is less intensive and was observed at temperatures up to $\sim 15$ K.  Modes E and F are relatively week, but still can be  detected  at lowest available  temperature, 1.4 K.

Using exact diagonalization (ED) calculations on a
22-spin cluster for an extended Kitaev-Heisenberg model, Yadav
\emph{et~al.} \cite{Yadav} predicted the presence of a field-induced
gapped quantum spin liquid state.  In addition to the Kitaev coupling
$K=-5.6$~meV,  in-plane $g$ factor $g_{ab}=2.51$, and isotropic
Heisenberg exchange interactions between nearest-, second-nearest-,
and third-nearest-neighbor sites, $J_1=1.2$~meV, $J_2=J_3=0.25$~meV,
the model includes nearest-neighbor symmetric anisotropic exchange
constants $\Gamma_{xy}=-1.2$~meV and
$\Gamma_{zx}=-\Gamma_{yz}=-0.7$~meV.  For the given set of parameters,
a spin liquid state was predicted  to exist in $\alpha$-RuCl$_3$  between approximately 11.5
and 14~T~\cite{Yadav}.  Notably, the crossover from the  spin-liquid to a spin-polarized phase should be accompanied by a pronounced dip
in the excitation energy at about 15 T  \cite{Yadav}. No indication of such a dip in
magnetic fields up to 16~T (at least for the chosen direction of the applied magnetic field) has been reveled in our experiments. 

Baek et al. ~\cite{Baek}  performed ED calculations  for the regular Kitaev-Heisenberg model on a 24-spin cluster, without symmetric anisotropic coupling and  assuming $K=-10.0$ meV, $g_{ab}=2.4$, $J_1=2.0$~meV, $J_2=J_3=0.5$~meV. The calculations were done for  intensities integrated over a broad momentum range and revealed a very rich excitation spectrum.

Recently, Winter~\emph{et al.} \cite{Winter_2} have reported another  ED study of $\alpha$-RuCl$_3$  for magnetic excitations  at the $\Gamma$ point  (which is the most common case for ESR) and 
including symmetric anisotropic coupling.  They considered a simplified $C_3$-symmetric four-parameter model,  assuming  $J_1=-0.5$ meV, $K_1=-5.0$ meV, $\Gamma_1=2.5$ meV,   $J_3=0.5$ meV. For fields $H>H_c$ the calculations showed a  large redistribution of spectral weight,  which can be attributed to the anisotropic frustration of the considered  model.    
Comparison of the calculation results for magnetic  field $H$ applied along the $a$ axis  (which corresponds to the [110] axis in our experiments) with the ESR data revealed a very good qualitative agreement.  The calculation results  are shown as color scale together with the experimental data in  Fig.~\ref{fig:FFD_b}. A  particular good agreement was obtained for the most intensive ESR mode C, suggesting that this mode corresponds to excitations at the $\Gamma$ point. The observation of a number of ESR modes, not accounted by the calculations \cite{Winter_2} for magnetic excitations  at the $\Gamma$ point, potentially suggests  the presence  of  exchange ESR modes (see the discussion above).   The corresponding calculations are in progress \cite{Winter_DM}.

Numerical calculations   \cite{Baek,Winter_2}  above $H_c$ predicted a rather complex excitation spectrum, consisting of a number of modes,  whose activation energy increases with increasing field.   A noticeable  high-field property of the detected   modes C and F   is their unusually large 
slope   $g_{ab}\mu_B\Delta S \approx 0.27$~meV/T (it is interesting, that the field dependence of mode F, the lowest mode observed in our ESR study in the $H>H_c$ region, matches that of the gap, extracted from the heat-transport experiments \cite{Hentric}). The remarkably large slope  might imply the presence of  ESR transitions with  $\Delta S \approx 2$   
(contrary to  $\Delta S = 1$, expected for elementary one-particle excitations in simple $S=1/2$ systems). 
This observation   (together with results of the ED calculations, evident of a number of excitation modes, split off from the higher-energy continuum),  suggests  that the spin dynamics of $\alpha$-RuCl$_3$   has an emergent multiparticle nature. To the best of our knowledge, such unconventional discrete (bound-state-like) multiparticle excitation spectrum   has never  been  previously observed in magnetic systems with a honeycomb lattice. One can speculate that strong ferromagnetic Kitaev coupling $K$ may facilitate the formation of two-magnon bound states that are strongly split down from the two-magnon continuum. However, at present stage, it is unclear
whether observed multiparticle excitations correspond to bound magnons or to bound Majorana spinons (and, in the latter case, whether the spinons are
confined or simply bound). We hope that our data will stimulate further theoretical studies of the unusual spin dynamics in $\alpha$-RuCl$_3$, in particular, in its high-field phase.

In conclusion, we performed comprehensive  high-field ESR studies of
$\alpha$-RuCl$_3$, an   anisotropic spin system
with a honeycomb structure  that is considered a top candidate for exhibiting Kitaev's spin-liquid physics. Our  experiments  revealed  the presence of  two soft AFMR modes in the zigzag-ordered phase and uncovered the 
rather complex  spin dynamics in the field-induced quantum paramagnetic state, characterized by emergent   multiparticle excitations.   Our observations can have  a broader impact, suggesting  that honeycomb-lattice magnets might  serve as an excellent playground to  study unconventional  many-body quantum  processes in condensed matter, including, e.g.,  field-induced condensation of bound states (with a potential realization of the spin-nematic order), Efimov effect, etc.

{\textit {Note:} Upon finalizing this manuscript, we became aware  of two other high-field spectroscopy  studies of $\alpha$-RuCl$_3$,  by  Wang et al. \cite{Zhe} and Wellm \cite{Wellm}.  Apart from the low-frequency AFMR mode   and a number of excitations in the field-induced disordered phase (partially similar to that observed by us),  a signature of a broad excitation continuum,  an  indication  of the unconventional multiparticle spin dynamics in $\alpha$-RuCl$_3$,   was revealed.

This work was partially supported by Deutsche Forschungsgemeinschaft
(project ZV 6/2-2) and by the HLD at HZDR, member of the European Magnetic
Field Laboratory (EMFL). A.~Banerjee and S.~E.~Nagler were supported
by the Division of Scientific User Facilities, Basic Energy Sciences
US DOE, P.~Lampen-Kelley and D.~G. Mandrus by the Gordon and Betty
Moore Foundations EPiQS Initiative through Grant GBMF4416, J.-Q.~Yan
and C.~A.~Bridges by the U.S. Department of Energy, Office of Science,
Office of Basic Energy Sciences, Materials Sciences and Engineering
Division. We acknowledge discussions with S.~M.~Winter, K.~Riedl, R.~Valent\'{\i},
M.~Zhitomirsky, G.~Jackeli,
and S.~Nishimoto. We thank  Ch.~Balz for his help   orienting the samples.


\begin{thebibliography}{99}



\bibitem{Kitaev} A.~Kitaev, Ann. Phys. {\bf 321}, 2 (2006).
\bibitem{Chaloupka}  J.~Chaloupka, G.~Jackeli, and G.~Khaliullin, Phys. Rev. Lett. {\bf 105}, 027204  (2010).
\bibitem{Yadav} R.~Yadav, N.~A.~Bogdanov, V.~M.~Katukuri, S.~Nishimoto, J.~van~den~Brink, and L.~Hozoi,
Sci. Rep. {\bf 6}, 37925 (2016).
\bibitem{Petrova} O.~Petrova, P.~Mellado, and O.~Tchernyshyov,  Phys. Rev. B {\bf 88}, 140405 (2013).
\bibitem{Baskaran} G.~Baskaran, S.~Mandal, and R.~Shankar,  Phys. Rev. Lett. {\bf 98}, 247201 (2007).
\bibitem{Knolle} J.~Knolle, D.~L.~Kovrizhin, J.~T.~Chalker, and R.~Moessner,  Phys. Rev. Lett. {\bf 112}, 207203 (2014).
\bibitem{Janssen} L.~Janssen, E.~C. Andrade, and M. Vojta, Phys. Rev. Lett.  {\bf 117}, 277202 (2016).
\bibitem{Do} S.-H.~Do, S.-Y. Park, J. Yoshitake, J.~Nasu, Y.~Motome, Y.-S.~Kwon, D.~T.~Adroja, D.~J. ~Voneshen, K.~Kim, T.-H.~Jang, , J.-H.~Park,
K.-Y.~ Choi, and S.~Ji, Nat. Phys. {\bf 13}, 1079 (2017).
\bibitem{Jackeli} G.~Jackeli and G.~Khaliullin, Phys. Rev. Lett. {\bf 102}, 017205 (2009).
\bibitem{Singh} Y.~Singh, S.~Manni, J.~Reuther, T.~Berlijn, R.~Thomale, W.~Ku, S.~Trebst, and  P.~ Gegenwart, Phys. Rev. Lett. {\bf 108}, 127203 (2012).
\bibitem{Takayama} T.~Takayama, A.~Kato, R.~Dinnebier, J.~Nuss, H.~Kono, L.~S.~I.~Veiga, G.~Fabbris, D.~Haskel, and H.~ Takagi, Phys. Rev. Lett. {\bf 114}, 077202 (2015).
\bibitem{Kubota} Y.~Kubota, H.~Tanaka, T.~Ono, Y.~Narumi, and K.~ Kindo, Phys. Rev. B {\bf 91}, 094422 (2015).
\bibitem{Johnson} R.~D.~Johnson, S.~C.~Williams, A.~A.~Haghighirad, J.~Singleton, V.~Zapf, P.~Manuel, I.~I.~Mazin, Y.~Li, H.~O.~Jeschke, R~Valent\'{\i}, and R. Coldea, Phys. Rev. B {\bf 92}, 235119 (2015).
\bibitem{Banerjee} A.~Banerjee, C.~A.~Bridges, J.-Q. ~Yan, A.~A.~Aczel, L.~Li, M.~B.~Stone, G.~E.~Granroth, M.~D.~Lumsden, Y.~Yiu, J.~Knolle, S.~Bhattacharjee, D.~L.~Kovrizhin, R.~Moessner, D.~A.~Tennant, D.~G.~Mandrus,  and S.~E.~Nagler, Nat. Matter  {\bf 15}, 733 (2016).
\bibitem{Ran} K.~ Ran, J.~Wang, W.~Wang, Z.-Y.~ Dong, X.~Ren,  S.~Bao,  S.~Li,  Z.~Ma,  Y.~Gan, Y.~Zhang,  J.~ T. ~Park, G.~ Deng,  S.~Danilkin, S.-L.~Yu, J.-X. ~Li,  and J.~ Wen, Phys. Rev. Lett. {\bf  118}, 107203 (2017).
\bibitem{Leahy} I.~A.~Leahy, Ch.~A.~Pocs, P.~E.~Siegfried, D.~Graf, S.-H.~Do, K.-Y.~Choi, B.~Normand, and M.~Lee, Phys. Rev. Lett. {\bf 118}, 187203 (2017).
\bibitem{Rem_FSPS} Strictly speaking, due to the presence of the Kitaev coupling,  a full spin polarization can be achieved  only at infinitely large magnetic field.
\bibitem{Baek} S.-H.~Baek, S.-H.~Do, K.-Y.~Choi, Y.~S.~Kwon, A.U.B.~Wolter, S.~Nishimoto, J.~van~den~Brink, and B.~B\"{u}chner, Phys. Rev. Lett. {\bf 119},  037201 (2017). 
\bibitem{Sears} J.~A.~Sears, Y.~Zhao, Z.~Xu, J.~W.~Lynn,  and Y.-J.~Kim, Phys. Rev. B {\bf 95},  180411 (2017).
\bibitem{Hentric} R.~Hentrich, A.~U.~B.~Wolter, X.~Zotos, W.~Brenig, D.~Nowak, A.~Isaeva, T.~Doert, A.~Banerjee, P.~Lampen-Kelley, D.~G. Mandrus, S.~E.~Nagler, J.~ Sears, Y.-J.~Kim, B.~B\"{u}chner, and Ch.~Hess, arXiv:1703.08623.
\bibitem{Wolter} A.~U.~B. Wolter, L.~T.~Corredor, L.~Janssen, K.~Nenkov, S.~Sch\"{o}necker, S.-H.~Do, K.-Y.~Choi, R.~Albrecht, J.~Hunger, T.~Doert, M.~Vojta, and B.~B\"{u}chner,  
Phys. Rev. B {\bf 96},  041405  (2017). 
\bibitem{Zvyagin_INSR} S.~A.~Zvyagin, J.~Krzystek, P.~H.~M.~van ~Loosdrecht, G.~Dhalenne, and A.~Revcolevschi, Physica B {\bf 346-347}, 1 (2004).
    \bibitem{rem_1} The formula $\nu = \sqrt{\Delta^2 - AH^2}$  (where $\nu$ is the excitation frequency, $\Delta$ is the zero-field gap, $A$ is a fit parameter, and $H$ is the applied magnetic field) was used to fit the low-field frequency-field dependences of the observed modes. The corresponding results are shown by dashed line in Fig.~\ref{fig:FFD_b}. 
\bibitem{Banerjee_2}  A.~Banerjee, J.~ Yan, J.~ Knolle, C.~A.~Bridges, M.~B.~ Stone, M.D.~ Lumsden, D.~G.~ Mandrus, D.~A.~ Tennant, R.~ Moessner, and S.~E.~ Nagler,  Science {\bf 356}, 1055 (2017).
	\bibitem{Little} A.~Little,~L.~Wu, P.~Lampen-Kelley, A.~Banerjee, S.~Pantankar, D.~Rees, C.~A.~Bridges, J.-Q.~Yan, D.~Mandrus, S.~E.~Nagler, and J.~Orenstein,  Phys. Rev. Lett. {\bf 119}, 227201 (2017).
	\bibitem{Winter} S.~M.~Winter, K.~Riedl, P.~A.~Maksimov, A.~L.~ Chernyshev,  A.~Honecker, and R~Valent\'{\i}, Nat. Comm. {\bf 8}, 1152 (2017). 
\bibitem{Bar} V.~G.~Bar'yakhtar, V.~V.~Yeremenko, V.~M.~Naumenko, Y.~G.~Pashkevich, V.~V.~Pishko, and V.~L.~Sobolev,  Sov. Phys. JETP {\bf 61}, 823 (19985). 
\bibitem{Sakai} T.~Sakai and  H.~Shiba, J. Phys. Soc. Japan {\bf 63}, 867 (1994).  
\bibitem{Zhe_BSCO} Z.~Wang, D.~Kamenskyi, O.~Cepas, M.~Schmidt, D.~L.~Quintero-Castro, A.~T.~M.~N.~Islam, B.~Lake, A.~A.~Aczel, H.~A.~Dabkowska, A.~B.~Dabkowski, G.~M.~Luke, Y.~Wan, A.~Loidl, M.~Ozerov, J.~Wosnitza, S.~A.~Zvyagin, and  J.~Deisenhofer Phys. Rev. B {\bf 89}, 174406 (2014).
\bibitem{Zvyagin_CUP} S.~A.~ Zvyagin, E.~ \v{C}i\v{z}m\'{a}r, M. ~Ozerov, J.~ Wosnitza, R.~ Feyerherm, S.~R.~ Manmana, and F.~ Mila,  Phys. Rev. B {\bf 83}, 060409(R) (2011). 
\bibitem{Winter_2} S.~M.~Winter, K.~Riedl, D.~Kaib, R.~Coldeau, and R.~Valent\'{\i}, arXiv:1707.08144.
\bibitem{Winter_DM} S.~M.~Winter, unpublished. 
\bibitem{Zhe} Z.~ Wang, S.~Reschke, D.~H\"{u}vonen, S.-H.~Do, K.-Y.~Choi, M.~Gensch, U.~Nagel, T.~R\~{o}\~{o}m, and A. Loidl, Phys. Rev. Lett. {\bf 119}, 227202 (2017). 
\bibitem{Wellm} C.~Wellm, J.~Zeisner, A.~Alfonsov, A.~ U.~ B.~ Wolter,  M.~Roslova,  A.~Isaeva,  T.~Doert,  M.~ Vojta, B.~B\"{u}chner,  and V. ~Kataev,   arXiv:1710.00670. 









\end{thebibliography}
\end{document}